\documentstyle[prb,aps,multicol,epsf,graphicx]{revtex}

\begin{document}
\draft
\title{Measurements of strongly-anisotropic $g$-factors for spins in
single quantum states}
\author{J.~R.~Petta and D. C.~Ralph}
\address{Laboratory of Atomic and Solid State Physics, Cornell
University, Ithaca, New York 14853}
\date{\today}
\maketitle

\begin{abstract}
We have measured the full angular dependence, as a function of the
direction of magnetic field, for the Zeeman splitting of individual
energy states in copper nanoparticles.  The $g$-factors for spin
splitting are highly anisotropic, with angular variations as large as
a factor of five.  The angular dependence fits well to ellipsoids.  Both
the principal-axis directions and $g$-factor magnitudes vary between
different energy levels within one nanoparticle.  The variations
agree quantitatively with random-matrix theory predictions
which incorporate spin-orbit coupling.
\end{abstract}

%\par
%\vspace{2.0cm}


\begin{multicols} {2}
\narrowtext
Strategies for manipulating electron spins are being pursued in many
fields\cite{wolf}, from spin-dependent tunneling\cite{moodera}, to
magnetic semiconductors\cite{ohno}, to quantum
computation\cite{loss}. An issue central to all of these fields is
that a spin is often not an independent variable. Interactions with
the environment, for instance coupling to orbital degrees of freedom,
can limit spin coherence times and alter responses to magnetic
fields. Here we study the effects of spin-orbit coupling on spins in
quantum dots by exploring in detail how the spin properties depend on
the direction of an applied magnetic field.  Specifically, we measure
the $g$-factor for spin Zeeman splitting, the sensitivity with which
an energy level shifts with magnetic field.  The angular dependence
of the $g$-factor has not previously been measured for quantum dots
formed from semiconductor 2-dimensional electron gases,
because out-of-plane magnetic-field components produce strong orbital
effects that make spin energies difficult to observe;
we avoid this problem by studying energy levels in copper
nanoparticles.  We find that the $g$-factors for each quantum state
can have striking anisotropies, varying by up to a factor of five
depending on field direction.  The full dependence on field angle is
ellipsoidal, as expected from the tensor properties of the
$g$-factor\cite{slichter,harriman}.  In addition, the
$g$-tensors exhibit large variations between different
energy eigenstates in the same nanoparticle, with differences both in
the direction of maximum field sensitivity and in magnitude. The
variations between levels suggest that the strong anisotropies
originate from the intrinsic quantum-mechanical fluctuations of
mesoscopic wavefunctions, an effect predicted by random-matrix
theories \cite{brouwer,matveev}.

In bulk systems and for impurities in solids, it is well understood
that spin-orbit coupling may cause $g$-factors to exhibit
anisotropies associated with directions of the crystal lattice
\cite{slichter,harriman}.
However, the results of the recent random-matrix-theory work
\cite{brouwer,matveev} suggest that a
different phenomenon may enhance the effects of spin-orbit coupling
in nanoscale grains.  The idea can be
understood by thinking about the wavefunction for an electron in a
metal nanoparticle, modeled as an electron in a box with an irregular
boundary.  The standing wave corresponding to the orbital amplitude
of the wavefunction will oscillate strongly and effectively randomly
as a function of position, and this complicated wave pattern will
vary from quantum state to quantum state.  In the presence of
spin-orbit coupling, the variations present in the orbital
wavefunction are reflected also in the spin properties of an
electronic state; in particular, they can cause the $g$-factor to
depend on the orientation of an applied magnetic field (${\bf B}$)
\cite{brouwer}. The most general dependence on the field direction
permitted by symmetry for a spin-$1/2$ particle allows the $g$-factor
$g_{\mu}$ for an energy level $\mu$ to be written in the form
\begin{equation}
g_{\mu}(\hat{\bf{B}})=(g_{1}^{2}B_{1}^{2}+
g_{2}^{2}B_{2}^{2}+g_{3}^{2}B_{3}^{2})^{1/2}/|\bf{B}|.
\end{equation}
where $g_{1}$, $g_{2}$, and $g_{3}$ are the $g$-factors along
mutually-orthogonal principal-axis directions (with the convention
$g_{1}$ $\ge$ $g_{2}$ $\ge$ $g_{3}$), and $B_{1}$, $B_{2}$, and
$B_{3}$ are the magnetic field components in these
directions\cite{harriman,brouwer}. Pictorially, this means that
magnitude of
the $g$-factor can be drawn as an ellipsoid.  In the absence of
spin-orbit coupling, the g-factor for spin splitting should be
isotropic. However, in the presence
of spin-orbit coupling, random-matrix theory predicts that the three
principal-axis g-factors for each level can differ strongly, and
the principal-axis directions can vary
randomly from energy level to energy level within the same
nanoparticle.

We have probed the spin properties of individual quantum states by
using electron tunneling to measure energy levels in single copper
nanoparticles.  We studied in detail two devices containing a
nanoparticle with approximately hemispherical shape and a diameter of
$\sim$15 nm, based on measured mean level spacings of 0.10 meV
(Cu\#1) and 0.12 meV (Cu$\#2$).  The nanoparticle is connected to
two aluminum electrodes by aluminum oxide tunnel junctions, as shown
schematically in the inset of Fig.\ 1. The procedures for fabricating
and characterizing the devices have been described in detail
previously\cite{ralls,petta}.
We use electron-beam lithography and reactive ion etching to create a
5-10 nm diameter hole in a silicon nitride membrane. A thick Al layer
is evaporated on one side of the sample (the bottom side in the inset
to Fig.\ 1) to fill the hole, and then this is oxidized in 50 mT
O$_2$ for three minutes to form an Al$_2$O$_3$ tunnel barrier. We
form the Cu nanoparticles on top of the membrane by
self-assembly during the evaporation of 10 $\AA$ of Cu. The 
\begin{figure}
\begin{center}
\leavevmode
\hspace*{-0.55 cm}
\includegraphics[height=6.8 cm, width=8.25 cm]{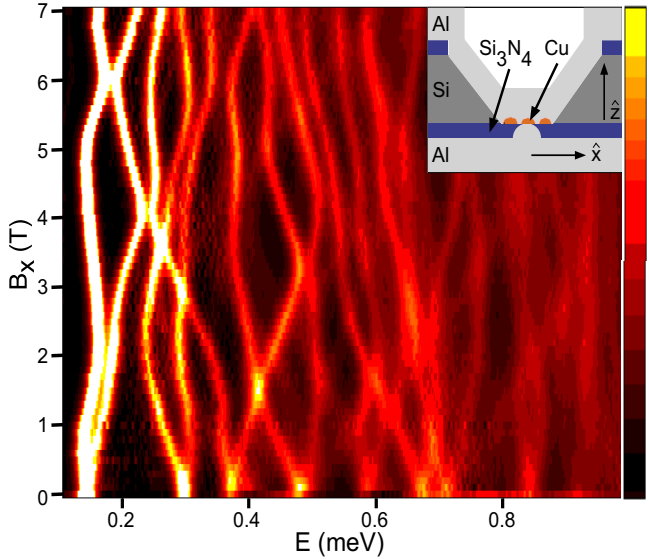}
\end{center}
\vspace{0.2 cm}
\caption{
\label{figure1}
Magnetic field dependence of quantum states in copper nanoparticle
Cu\#1. Main panel: A color-scale plot of the differential conductance
  ($dI$/$dV$) versus energy ($E$) as a function of magnetic field ($B_{x}$).
Yellow corresponds to 2 mS, and black to zero.  Values between 2 mS and
the maximum conductance of 3 mS are set to white.  The energy is equal to
the value of source-drain voltage multiplied by the capacitance 
factor$^{12}$ $eC_{1}/(C_{1}+C_{2}) = 0.66e$. The complex evolution of the
levels with $B_{x}$ is due to spin-orbit coupling between spin-up and
spin-down quantum states. The inset shows a sample schematic. A small hole
in a $Si_{3}$$N_{4}$ membrane is used to make electrical contact to a
single copper nanoparticle. $Al_{2}$$O_{3}$ tunnel junctions (not shown)
lie between the nanoparticle and aluminum electrodes.
}
\end{figure}
\noindent
tunnel barrier on top of the nanoparticle is then made by depositing 11 
$\AA$ of
Al$_2$O$_3$ by electron-beam evaporation. Finally, we deposit a thick
layer of Al to form the top electrode.
We measure current-voltage ($I$-$V$) curves for electrons tunneling
from one electrode to the other via the nanoparticle.  The
measurements are conducted using a dilution refrigerator with an
electron base temperature of approximately 40 mK.  Once $V$ exceeds a
threshold associated with the Coulomb-blockade effect\cite{averin},
tunneling via the discrete quantum states in the nanoparticle causes
the $I$-$V$ curve to take the form of a sequence of small steps,
allowing a direct measurement of the spectrum of ``electron-in-a-box"
states\cite{ralph}.

In Fig.\ 1 we plot, using a color-scale, differential conductance
($dI/dV$) versus energy ($E$) curves for different values of magnetic
field in the x-direction (defined in the Fig.\ 1 inset),
showing how the energy levels in Cu\#1 evolve with $B_x$.  In this
figure, $dI/dV$ is obtained by numerical differentiating the $I$-$V$
curves, and the energy is determined by scaling the source-drain
voltage to correct for the capacitive division of voltage across the
two tunnel junctions in the device\cite{ralph}.  As has been observed
previously\cite{petta,ralph}, the energy levels exhibit two-fold
Kramers' degeneracy at $B$=$0$ associated with the electron spin, and
they undergo Zeeman splitting as a function of increasing $B$.  We
determine the $g$-factor from the magnitude of the energy splitting
between Zeeman-split states, in the small-$B$ linear regime:
$g$=$\Delta$$E(\bf{B})$/($\mu_{B}$$|\bf{B}|$). The data in Fig.\ 1
exhibit different values of the $g$-factor for different quantum
states and a rich variety of avoided level crossings as a function of
increasing $B$.  The role of spin-orbit coupling in producing these
two effects has been discussed
previously\cite{petta,salinas,davidovic}.

Our focus in this letter is the unexplored question of how
the $g$-factors
for the individual quantum-dot states depend on the direction of the applied
magnetic field.  We control the field magnitude and direction using a
3-coil magnet which can produce 1 T in any direction, or up to 7 T
along one axis\cite{ami}.  We first verified that the tunneling
spectra were identical for $\bf{B}$ and $\bf{-B}$, as expected due to
time-reversal symmetry.  Thereafter, we took detailed measurements
only for $B_{x}$$>$0.  For each sample, we measured the Zeeman
splitting as a function of $|\bf{B}|$ along 62 different directions
selected to span this hemisphere\cite{hardin}.  The values of the
$g$-factors showed strong anisotropies, described well by an
ellipsoid for each energy level, as demonstrated below.  After
fitting for the directions of the three principal axes of the
$g$-tensor ($\hat{h}_{1}$, $\hat{h}_{2}$, $\hat{h}_{3}$) for each
state, we took detailed field scans along these directions to
determine precisely the principal-axis $g$-factors, and along arcs of
constant field to carefully measure the variation of $g$ with field
direction. The long measurement time (1-2 weeks for each sample)
requires exceedingly good sample stability.

The directional dependence of $g$ for the first quantum state in
sample Cu\#1 is shown in Fig.\ 2.  In Fig.\ 2(a), $dI/dV$ vs.\ $E$ is
plotted as a function of the angle of the magnetic field (with
magnitude 700 mT), rotating the field vector around the $\hat{h}_{3}$
principal axis for the first quantum state from an initial
orientation along $\hat{h}_{2}$, where $g$ = 0.51$\pm$0.05, to a
final orientation along $\hat{h}_{1}$, where $g$ = 0.86$\pm$0.08. The
fitted $g$-factors are in good agreement with the form $g(\alpha) =
\sqrt{g_{1}^2{\rm sin}^{2}(\alpha)+g_{2}^2{\rm cos}^{2}(\alpha)}$
expected for an ellipsoid. The data corresponding to a rotation from
$\hat{h}_{3}$ (where $g$ = 0.38$\pm$0.05) to $\hat{h}_{1}$ about the
$\hat{h}_{2}$ axis are shown in plots (c-d), and for rotation from
$\hat{h}_{3}$ to $\hat{h}_{2}$ in (e-f). The $g$-factors along the
$\hat{h}_{1}$ and $\hat{h}_{3}$ axes for this quantum state differ by
a factor of 2.3. Fig.~2(a) also displays the existence of
energy-level to energy-level variations in spin properties.
As $\alpha$ increases from 0 to $\pi$/2, the splitting of the first
quantum state grows while the splitting of the fourth Zeeman-pair
decreases. The other electronic states in Fig.\ 2 show more
complicated behavior due to the alignment of their principal axes in
other directions in space. We obtain good fits to ellipsoids for
all of the measured states, with the exception of those that undergo
an avoided crossing with a neighboring state at a very low value of 
field, so that the $g$-factor cannot be determined accurately (states
6 and 7 of Cu\#1 and states 1, 6, and 7 of Cu\#2).  The fitted parameters
of the measured $g$-factor ellipsoids are listed in Table
1.  There are large variations between different\linebreak
\begin{figure}
\begin{center}
\leavevmode
\includegraphics[height=11 cm, width=8.5 cm]{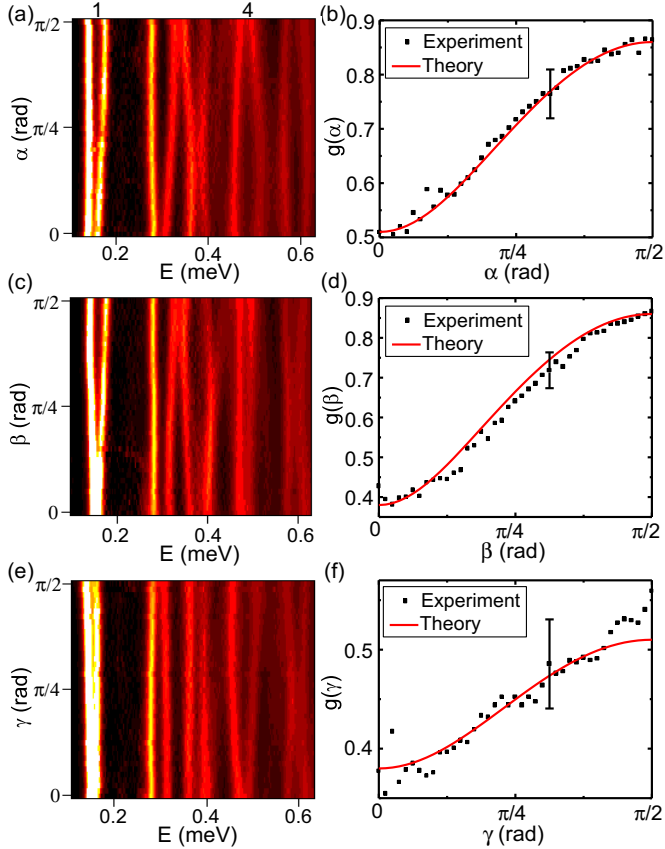}
\end{center}
\vspace{0.5 cm}
\caption{
\label{figure 2}
Dependence of Zeeman splitting on magnetic-field direction
in sample Cu\#1. (a), (c), (e) Differential conductance
($dI$/$dV$) as a function of energy ($E$) and the
angle of the magnetic field ($\alpha$, $\beta$, or $\gamma$,
respectively) for a fixed field magnitude of 700 mT. The
data in (a) correspond to a rotation of the field around the
$\hat{h}_{3}$ axis for the lowest-energy Zeeman-split
energy-level doublet from an initial direction along $\hat{h}_{2}$
to a final direction along $\hat{h}_{1}$, while plots (c) and (e)
correspond to rotations from $\hat{h}_{3}$ to $\hat{h}_{1}$
and $\hat{h}_{3}$ to $\hat{h}_{2}$, respectively.  The
color scale is identical to that used in Fig. 1. The g-factor of the
first Zeeman-split doublet is shown in (b), (d), and (f) as a function
of the rotation angles and is compared with the ellipsoidal dependence
expected from theory.}
\end{figure}
\noindent
quantum states in the same sample, both in the magnitude of the 
principal-axis
$g$-factors and in the direction of maximum field sensitivity.

Random matrix theory makes quantitative predictions about the
statistical distributions of the $g$-factors along each of the
principal axes, and how these distributions should depend on the
strength of the spin-orbit coupling parameter
($\lambda^{2}$$\propto$1$/$$\tau_{so}^2$). Within the theory of
\cite{brouwer}, the distributions of $g_{1}$, $g_{2}$, and $g_{3}$
may be generated for any $\lambda$, and the value of $\lambda$
corresponding to a particular nanoparticle is determined by matching
to the measured quantity
$\langle{g_{1}^2}\rangle$+$\langle{g_{2}^2}\rangle$+$\langle{g_{3}^2}\rangle$.
Here $\langle$$\rangle$ represents an average over energy eigenstates
in a particle.   We find $\lambda$=1.8$\pm$0.1 for
\vspace{0.3 cm}
\begin{table}
\caption{Principal-axis g-factors and principal-axis directions
for individual energy levels in samples
Cu\#1 and Cu\#2.  $g_{1}$,
$g_{2}$,
and $g_{3}$ are the principal-axis g-factors. $\theta_{1}$ and
$\phi_{1}$ are the spherical polar coordinates for the
$\hat{h}_{1}$ principal-axis direction while $\theta_{3}$ and
$\phi_{3}$ give the $\hat{h}_{3}$ direction. $\theta$ is measured
with respect to the $\hat{z}$ direction (perpendicular to the nitride
membrane), while $\phi$ is measured with respect to the x-axis. All
angles are in radians. $\hat{h}_{2}$ can be determined as $\hat{h}_{1}$
$\times$ $\hat{h}_{3}$. The g-factors of states 6 and
7 in Cu\#1 and 1, 6, and 7 in Cu\#2 could not be measured accurately
due to nonlinearities associated with avoided crossings between
closely-spaced states.}
\vspace{0.2 cm}
\begin{tabular}{cccccccc}
level&$g_{1}$&$g_{2}$&$g_{3}$&$\theta_{1}$&$\phi_{1}$&$\theta_{3}$&$\phi_{3}$
\\ \hline
Cu\#1/1&0.86$\pm$0.08&0.51$\pm$0.05&0.38$\pm$0.05&0.71&4.64&0.99&2.22
\\
Cu\#1/2&1.51$\pm$0.05&1.06$\pm$0.04&0.45$\pm$0.03&0.94&4.30&0.77&0.43
\\
Cu\#1/3&1.65$\pm$0.09&0.79$\pm$0.08&0.30$\pm$0.06&0.13&3.13&1.54&4.76
\\
Cu\#1/4&1.5$\pm$0.1&1.0$\pm$0.1&0.3$\pm$0.1&1.56&0.39&0.51&1.98   \\
Cu\#1/5&1.34$\pm$0.06&0.78$\pm$0.07&0.71$\pm$0.07&1.56&1.31&1.54&6.06
\\
Cu\#1/8&1.1$\pm$0.2&0.6$\pm$0.2&0.3$\pm$0.2&0.20&3.20&1.44&0.97    \\ \\
Cu\#2/2&1.3$\pm$0.2&1.2$\pm$0.2&0.9$\pm$0.1&0.46&5.72&1.24&3.37    \\
Cu\#2/3&1.8$\pm$0.2&1.4$\pm$0.2&1.2$\pm$0.2&1.49&5.85&0.08&2.33  \\
Cu\#2/4&1.6$\pm$0.3&0.9$\pm$0.2&0.6$\pm$0.1&1.51&1.89&0.98&0.28  \\
Cu\#2/5&1.9$\pm$0.1&1.0$\pm$0.2&0.5$\pm$0.1&1.39&5.02&0.20&2.42  \\
Cu\#2/8&1.3$\pm$0.2&1.2$\pm$0.1&0.9$\pm$0.1&0.74&1.21&0.83&4.52  \\
Cu\#2/9&1.9$\pm$0.1&1.4$\pm$0.1&1.2$\pm$0.2&0.65&3.96&0.92&0.78  \\
\end{tabular}
\end{table}
\noindent
Cu\#1 and $\lambda$=1.1$\pm$0.1 for Cu\#2.  Because spin-orbit 
coupling is
thought to be dominated by scattering from defects and surfaces, it
is not surprising that we find slightly different values of $\lambda$
for two particles made from the same material \cite{petta}.  Given
these $\lambda$, the average values for the $g$-factors along the
principal axes, calculated assuming that the contributions to Zeeman
splitting from orbital moments can be neglected relative to spin
moments \cite{brouwer,petta} are predicted to be
$\langle{g_{1}}\rangle$=1.25, $\langle{g_{2}}\rangle$=0.76, and
$\langle{g_{3}}\rangle$= 0.52 for Cu\#1, and for Cu\#2
$\langle{g_{1}}\rangle$= 1.59,   $\langle{g_{2}}\rangle$= 1.12, and
$\langle{g_{3}}\rangle$= 0.96. The experimental averages calculated
from Table 1 are   $\langle{g_{1}}\rangle$= 1.3$\pm$0.3,
$\langle{g_{2}}\rangle$= 0.8$\pm$0.2, and   $\langle{g_{3}}\rangle$=
0.4$\pm$0.2 for Cu\#1 and   $\langle{g_{1}}\rangle$= 1.6$\pm$0.3,
$\langle{g_{2}}\rangle$= 1.2$\pm$0.2, and   $\langle{g_{3}}\rangle$=
0.9$\pm$0.3 for Cu\#2. Therefore, using one fitting parameter
($\lambda$) per sample, our results are in excellent agreement with
the predictions.
\\
\indent
The distribution of principal-axis directions can provide insight as to
whether the axes are preferentially aligned due to crystalline axes or
sample shape, or randomly oriented, as would be expected for
spins coupling to highly-oscillatory electron-in-a-box orbital
wavefunctions \cite{brouwer}. We tested whether the principal
axes that we measure show a tendency toward any preferred direction
by plotting the axes on polar plots (Fig.\ 3).  The angles $\theta$
and $\phi$ are spherical polar coordinates, with $\theta$ measured
relative to $\hat{z}$, the unit vector perpendicular to the plane of
the sample's silicon nitride membrane.  The circles of constant
$\theta$ are spaced \pagebreak so that equal areas \linebreak 
\begin{figure}
\vspace{-2.45 cm}
\begin{center}
\leavevmode
\includegraphics[height=6.5 cm, width=8.5 cm]{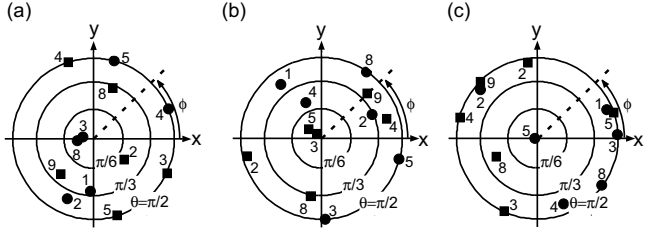}
\end{center}
\vspace{-1.5 cm}
\caption{
\label{figure 3}
Distributions of principal-axis directions. (a), (b), (c). Data points
represent the $\hat{h}_{1}$, $\hat{h}_{2}$, and
$\hat{h}_{3}$ principal-axis
directions, respectively, with circles for sample Cu\#1 and squares
for Cu\#2.
The numbers identify the energy levels (see Table I), in order of
increasing excitation energy.
$\theta$ and $\phi$ are spherical polar coordinates, with
$\theta$ measured relative to $\hat{z}$, the direction perpendicular to
the plane of the silicon nitride membrane.  The radii of the circles of
constant $\theta$ are scaled so that the areas inside the circles are
proportional to the solid angle of the sphere with polar angle
less than $\theta$.}
\end{figure}
\noindent
of solid angle on the
hemisphere are mapped to equal areas in Fig.\ 3. The principal 
axes
for different quantum states clearly do vary widely, and there does
not appear to be any clustering in particular directions.

In summary, we have made the first measurements of the angular
dependence of the $g$-factors for spin Zeeman splitting for
individual quantum states in quantum
dots. The results verify the predictions that
spin-orbit coupling can cause the spin properties of quantum states
to depend sensitively on the direction of an applied magnetic field.
When we compare the $g$-factors between different quantum states in
the same nanoparticle, there are large variations both in angular dependence
as well as magnitude, supporting the picture that the strong
anisotropies arise from the intrinsic random variations of
quantum-mechanical wavefunctions.

We thank Edgar Bonet, Mandar Deshmukh, and Eric Smith for help with
the experimental setup, and Xavier Waintal, Shaffique Adam, and Piet
Brouwer for discussions.  This work was supported by the NSF, through
DMR-0071631 and the use of the National Nanofabrication Users
Network.  Support was also provided by the Packard Foundation, ARO
(DAAD19-01-1-0541), and a DoEd GAANN fellowship.

\end{multicols}
\end{document}